\documentclass{amsart}

\usepackage{graphicx}

\input{epsf}

\theoremstyle{definition}

\theoremstyle{remark}

\numberwithin{equation}{section}

\newcommand{\BS}{\boldsymbol \sigma}
\newcommand{\Bgrad}{\boldsymbol \nabla}
\newcommand{\sign}{{\rm\,  sgn}}

\newcommand{\bb}{\begin{equation}}
\newcommand{\ee}{\end{equation}}
\newcommand{\eqb}{\begin{eqnarray}}
\newcommand{\eqf}{\end{eqnarray}}

\begin{document}

\title[]{Graphene and non-Abelian quantization}%



\author{H.\ Falomir$^a$, J.\ Gamboa$^{b,c}$, M.\ Loewe$^{b,d}$ and M.\ Nieto$^e$}%

\address{$a$: IFLP, CONICET - Departamento de F\'{\i}sica, Fac.\ de Ciencias Exactas de la UNLP, C.C. 67, (1900) La PLata, Argentina.}%
\address{$b$: Facultad de F\'{\i}sica, Pontificia Universidad Cat\'{o}lica de Chile, Casilla 306, San\-tia\-go 22, Chile.}%
\address{$c$: Departamento de F\'{\i}sica, Universidad de Santiago de Chile, Casilla 307, San\-tia\-go, Chile.}%
\address{$d$: Centre for Theoretical and Mathematical Physics, University of Cape Town, Rondebosch 770, South Africa.}%
\address{$e$: Departamento de F\'{\i}sica, Fac.\ de Ciencias Exactas de la UNLP, C.C. 67, (1900) La PLata, Argentina.}%

\email{falomir@fisica.unlp.edu.ar; jgamboa55@gmail.com; mloewe@fis.puc.cl; mnieto@fisica.unlp.edu.ar;}%


\begin{abstract}
In this article we employ a simple nonrelativistic model to describe the low energy excitation of graphene. The model is based on a deformation of the Heisenberg algebra which makes the commutator of momenta proportional to the pseudo-spin. We solve the Landau problem for the resulting Hamiltonian which reduces, in the large mass limit while keeping fixed the Fermi velocity, to the usual linear one employed to describe these excitations as massless Dirac fermions. This model, extended to negative mass, allows to reproduce the leading terms in the low energy expansion of the dispersion relation for both nearest and next-to-nearest neighbor interactions. Taking into account the contributions of both Dirac points, the resulting Hall conductivity, evaluated with a $\zeta$-function approach, is consistent with the anomalous integer quantum Hall effect found in graphene. Moreover, when considered in first order perturbation theory, it is shown that the next-to-leading term in the interaction between nearest neighbor produces no modifications in the spectrum of the model while an electric field perpendicular to the magnetic field produces just a rigid shift of this spectrum.


\end{abstract}

\maketitle

\setcounter{figure}{0}

\section{Introduction}\label{introduccion}

Several theoretical ideas have produced a fruitful exchange of interpretations and methods between fundamental physics and condensed matter theory, such as spontaneously broken symmetry \cite{1,2,3,4} or renormalization group methods \cite{5,6,7,8,9}.

The recent experimental construction of graphene \cite{Novo-2004} opens a new connection between condensed matter and quantum field theory, since its low energy excitations can be represented as massless planar fermions and described by means of a (pseudo) relativistic theory.

Gaphene (See \cite{Castro,Vozmediano} and references therein) is a two-dimensional, one-atom-thick, allotrope of carbon which has attracted great attention in the last years. The carbon atoms are arranged on a honeycomb structure made out of hexagons, a structure with a great versatility. Carbon nanotubes, for example, can be obtained by rolling the graphene plane along a given direction and reconnecting the carbon bonds at the boundaries, giving rise to essentially one-dimensional object. Also, the replacement of an hexagon by a pentagon in this lattice introduces a positive curvature defect; this allows to wrap-up graphene to give fullerenes, molecules where carbon atoms are arranged on spherical structures. Notice that the accumulation of graphene layers, weakly coupled by van der Waals forces, constitute the well-known graphite.

This peculiar material has been theoretically predicted by Semenoff in 1984 \cite{Semenoff} (See also \cite{Wallace}) and experimentally produced in the lab in 2004 \cite{Novo-2004}.

The electronic properties of graphene are the result of the $sp^2$ hybridization between one $s$ and two $p$ orbitals, which leads to a trigonal planar lattice with a $\sigma$ bond between carbon atoms separated by $1.42$ Angstrom. This filled band gives the lattice its robustness. The third $p$ orbital of the carbon atom, oriented perpendicularly to the plane of the graphene, gives rise to the $\pi$ band through covalent bounds with neighboring atoms. Since this $p$ orbital contributes with only one electron, the $\pi$ band is half-filled in neutral graphene.

The low energy excitations of graphene accept a description as states of chiral massless Dirac fermions with a pseudo-relativistic linear dispersion relation, in which the speed of light $c$ is replaced by the \emph{Fermi velocity}, $v_F\approx 10^{-3} c$. Then, the Lagrangian describing these low energy states in the presence of an Electromagnetic field is similar to that of QED for massless fermions, shearing therefore some of its peculiarities. In particular, when a magnetic field is applied perpendicularly to the plane of graphene, an anomalous integer quantum Hall effect \cite{Gusynin-2005,Peres-2006} takes place, which has been experimentally measured \cite{Novo-2005,Zhang}.

\medskip

It is not the goal of the present paper to discuss the physics of graphene from first principles. Rather, we will consider a simple effective non-relativistic Hamiltonian, suggested by a particular deformation of the Heisenberg algebra (non-commutativity of momenta, consistent with the introduction of an external constant non-Abelian magnetic field), which could be useful to describe the low energy excitations produced by the dominant nearest and next-to-nearest neighbor interactions in graphene.

\medskip

The non-commutativity of spacetime is an old idea \cite{Snyder}, the first example of which was probably discussed by Landau in 1930 \cite{Landau}. It has been revived in recent years within the context of string theory \cite{Douglas} and since then, non-commutative field theories have attracted much attention in various fields such as Mathematics, Theoretical Physics ,  \cite{Connes1,Connes2,Witten,Seiberg}, and Phenomenology \cite{Hinchliffe}.

The breaking of commutativity of the position operators and the representations of the algebra of the non-commutative space-time coordinates has been studied in \cite{Doplicher}. The  non-commutativity in the momenta algebra we are interested in can be related to the deformation quantization of Poissonian structures developed in  \cite{Rie} and considered as a kind of magnetic quantization \cite{Ift,Man}.

All these researches have stimulated the construction of new models in quantum mechanics \cite{NCQM}, which have opened new routes to explore, for example in superconductivity [6]. Also, a massless Dirac-like Hamiltonian in a generalized noncommutative space and its relation with graphene has been considered in \cite{Dayi}.

Recently, some models based on a kind of nonstandard deformation of the Heisenberg algebra, which can be realized by shifting the dynamical variables with the spin, have been studied in \cite{SpinNC1,SpinNC2}. In the following, we will consider a similar deformation, but concentrated in the commutators among momenta, which can be interpreted as the introduction of a constant non-Abelian magnetic field \cite{weiss}.

\medskip

In the next Section we present the model and derive the Hamiltonian. In Section \ref{free-case} we study the free case and in Section \ref{Constant-magnetic} we introduce a constant magnetic field and solve the corresponding Landau problem. In Section \ref{relation-graphene} we apply our results to describe the low energy states associated to the leading nearest and next-to-nearest interactions in graphene.

In Section \ref{Hall-conductivity} we evaluate the associated Hall conductivity employing a $\zeta$-function approach, finding that the result is consistent with the anomalous integer quantum Hall effect present in this material. In Section \ref{perturbation} we are also able to show that the next-to-leading (quadratic) term in the nearest neighbor interaction does not change the spectrum at first order in perturbation theory. In Section \ref{crossed} we comment on the case where crossed electric and magnetic fields are present and, finally, in Section \ref{Conclusions} we establish our conclusions.

At the end of the paper, Appendix \ref{Ec-Mov} is dedicated to the Lagrangian of the model and studies its symmetry, conserved current and the generating functional of Green's functions. It also comments on the weak field and gradient expansion of the generating functional, its relation with the Hall conductivity and the topological considerations involved.

\section{Deformation of the Heisenberg algebra} \label{NC-space}

We consider particles leaving on a plane whose dynamical variables satisfy the deformed Heisenberg algebra given by
\begin{equation}\label{Ha}
      \left[X_i , X_j\right]=0\,,
    \quad
      \left[ X_i , P_j\right] = \imath  \delta_{i j} \,,
    \quad
      \left[P_i , P_j\right] = 2 \imath \theta^2 \epsilon_{i j 3} \sigma_3 \,, \quad i,j=1,2\,,
\end{equation}
where the momenta commutator is proportional to the \emph{pseudospin} $\sigma_3$ and $\theta$ is a parameter with dimensions of momentum (For convenience, we take $\hbar=1$ and return to full units when necesary).

Notice that these particles are described by wave functions with two components, $\psi=\left(\begin{array}{c}
\varphi \\
\chi \\
\end{array}
\right)\in \mathbf{L}_2\left(\mathbb{R}^2\right) \otimes \mathbb{C}^2$, and these operators have the structure of $2\times 2$ matrices on $ \mathbb{C}^2$.

The deformed algebra in Eq.\ (\ref{Ha}) can be realized by defining
\begin{equation}\label{var}
    X_i:=x_i \otimes \mathbf{1}_2 \qquad P_i:=p_i \otimes \mathbf{1}_2+\theta\,\mathbf{1}_{\mathbf{L}_2}  \otimes\sigma_i\,,
\end{equation}
with $x_i\,, i=1,2$ the usual (commutative) coordinates on the plane and $p_i=-\imath\, \partial_i$, operators on $\mathbf{L}_2\left(\mathbb{R}^{2}\right)$, and $\sigma_i\,, i=1,2$ the first two Pauli matrices. We will also write $\BS:=\left(\mathbf{1}_{\mathbf{L}_2}  \otimes\sigma_1,\mathbf{1}_{\mathbf{L}_2}  \otimes\sigma_2\right)$. For notational convenience, from now on we will avoid the explicit indication of the symbol $\otimes$, which can lead to no confusion.

\smallskip

Our aim is to consider the direct generalization of the Hamiltonian of a (nonrelativistic) particle of charge $e$ and mass $m$, minimally  coupled to an external (2+1-dimensional) electromagnetic field, $\left\{A_0,\mathbf{A}:=\left(A_1,A_2\right)\right\}$, which is constructed by means of the replacements $x_i\rightarrow X_i$, $p_i\rightarrow P_i$.

For the time being, we make $A_0=0$. Then,
\begin{equation}\label{Ham}
    H_0=\frac{\left( \mathbf{p} -e \mathbf{A} \right)^{2}}{2 m}
    \quad \rightarrow \quad
    H_\theta=\frac{\left( \mathbf{p}-e \mathbf{A}+\theta \BS \right)^{2}}{2 m}\,,
\end{equation}
where the electromagnetic field is taken in the Coulomb gauge, $\Bgrad\cdot \mathbf{A}=0$.

The Hamiltonian can also be written as
\begin{equation}\label{Hma-1}
    H=\frac{\left( \mathbf{p}-e \mathbf{A} \right)^{2}}{2 m}
    + v_F \, \BS \cdot \left( \mathbf{p}-e \mathbf{A} \right)\,,
\end{equation}
where we have defined the \emph{Fermi velocity}
\begin{equation}\label{vf}
    v_F:=\frac{\theta}{m}
\end{equation}
and subtracted the constant $\theta^2/m$.

In the $m\rightarrow\infty$ limit, with fixed $v_F$, the resulting linear Hamiltonian is appropriate to describe the conducting effective particles in graphene around the \emph{Fermi points} \cite{Semenoff,Castro,Vozmediano,Gusynin-2005}, which justify our proposal. Notice that this limit does not correspond to a small but rather to a large deformation of the commutator $\left[P_1,P_2\right]$.

\medskip

Notice also that the modification of the Hamiltonian in Eq.\ (\ref{Hma-1}) can also be interpreted as the introduction of an $SU(2)$ non-Abelian constant and uniform \lq \lq magnetic" field $\sim \theta^2$. Indeed, the $SU(2)$ transformations relate both components of the wave functions, and the commutator of \emph{covariant} derivatives gives
\begin{equation}\label{foot1}
    \begin{array}{c}\displaystyle
      \left[p_1-e A_1 +\theta \sigma_1, p_2-e A_2 +\theta \sigma_2 \right]=
      \imath e \left(\partial_1 A_2-\partial_2 A_1\right)+2 \imath\theta^2 \sigma_3 \,.
    \end{array}
\end{equation}
In this sense, if we take a constant magnetic field $B=\left(\partial_1 A_2-\partial_2 A_1\right)$, the system we are considering is a kind of non-Abelian version of the Landau problem \cite{weiss}.

In Appendix \ref{Ec-Mov} we describe the Lagrangian of this model, study its symmetry and conserved current and discuss the relation of the  asymptotic expansion  of its generating functional with the Hall conductivity.

\section{The free case}\label{free-case}

In this Section we consider the \emph{free} case, with $\mathbf{A}=0$. The Hamiltonian reduces to
\begin{equation}\label{Hma-2}
    H=\frac{\mathbf{p}^{2}}{2 m}
    + v_F \, \BS \cdot \mathbf{p}\,.
\end{equation}

We propose solutions of the form
\begin{equation}\label{eigen}
    \psi_{\mathbf{k}}(\mathbf{x})=e^{\imath \mathbf{k}\cdot \mathbf{x}} \chi(\mathbf{k})\,,
\end{equation}
with $\chi(\mathbf{k})\in \mathbb{C}^2$, which replaced in
\begin{equation}\label{autovalor}
    \left[H-\mathcal{E}(\mathbf{k})\right] \psi_{\mathbf{k}}(\mathbf{x}) = 0
\end{equation}
lead to
\begin{equation}\label{autovalor-1}
        \left\{\frac{\mathbf{k}^{2}}{2 m}+ v_F \, \BS \cdot \mathbf{k} -\mathcal{E}(\mathbf{k})\right\}\chi(\mathbf{k})=
    0\,.
\end{equation}
Nontrivial solutions require
\begin{equation}\label{autoval}
    {\rm det} \left\{\frac{\mathbf{k}^{2}}{2 m}+ v_F \, \BS \cdot \mathbf{k} -\mathcal{E}(\mathbf{k})\right\}=0\,,
\end{equation}
which gives
\begin{equation}\label{autovalor-2}
         {v_F}^2 \, {\mathbf{k}}^2 =
    \left[ \mathcal{E}(\mathbf{k})- \frac{\mathbf{k}^{2}}{2 m}
    \right]^2\,.
\end{equation}
\begin{figure} \label{dispersion-free.fig}
\epsffile{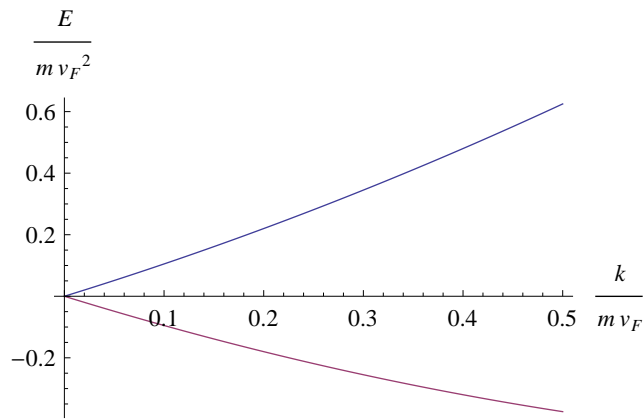}
\caption{The dispersion relations for the two branches of solutions for the free case.}
\end{figure}
Then, we get the following dispersion relation (approximately linear for small $\left| \mathbf{k} \right|$, see Figure \nolinebreak 1)
\begin{equation}\label{disp}
    \mathcal{E}(\mathbf{k})=\frac{\mathbf{k}^{2}}{2 m}
    \pm v_F \left| \mathbf{k} \right|
\end{equation}
which, replaced in Eq.\ (\ref{autovalor-1}) for $\mathbf{k}\neq 0$, shows that the pseudo-spinor $\chi(\mathbf{k})$ has definite pseudo-helicity,
\begin{equation}\label{disp-2}
    \left(\BS\cdot\mathbf{\hat{k}}\right)\chi_{\pm}(\mathbf{k})=\pm\, \chi_{\pm}(\mathbf{k})\,.
\end{equation}

On the other hand, for $\mathbf{k}=\boldsymbol{0}$ the two linearly independent vectors $\left(\begin{array}{c} 1 \\ 0 \\ \end{array} \right)$ and $\left(\begin{array}{c} 0 \\ 1 \\ \end{array} \right)$ are just constants solutions with vanishing eigenvalue.

\medskip

The Hamiltonian in Eq.\ (\ref{Hma-2}) commutes with the \emph{effective angular momentum}, the generator of a $U(1)$ symmetry,
\begin{equation}\label{SU2-1}
    J:=-\imath \partial_\varphi + \frac{1}{2}\, \sigma_3\,, \quad \left[H , J\right]=0\,.
\end{equation}

Under a rotation on the plane, the wavefunction in Eq.\ (\ref{eigen}) changes into
\begin{equation}\label{SU2-2}
    \mathcal{U}(\vartheta) \psi(\mathbf{x}):=e^{\imath\, \frac{\vartheta}{2}\, \sigma_3} \psi\left({R(\vartheta)}^{-1} \mathbf{x} \right) = e^{\imath \mathbf{k}\cdot \left({R(\vartheta)}^{-1} \mathbf{x} \right)} e^{\imath\, \frac{\vartheta}{2}\, \sigma_3} \chi(\mathbf{k})\,.
\end{equation}

For $\vartheta=2\pi$ we have $R(2\pi)={\mathbf{1}}_3$ and $e^{\imath\pi \sigma_3}=-\mathbf{1}_2$. So, we get
\begin{equation}\label{SU2-3}
    \mathcal{U}(2\pi) \psi(\mathbf{x})=-\psi(\mathbf{x})\,.
\end{equation}
Therefore, these particles can be considered as fermions \cite{Castro}.

\section{Constant magnetic field perpendicular to the plane}\label{Constant-magnetic}

In this Section we consider the electromagnetic vector potential of a constant magnetic field orthogonal to the plane,
\begin{equation}\label{B1}
    \mathbf{A}= B x_1 \mathbf{\hat{e}}_2  \quad \Rightarrow \quad  \partial_1 A_2 - \partial_2 A_1=B \quad {\rm and}
    \quad \Bgrad\cdot \mathbf{A}=0\,.
\end{equation}

In this case the Hamiltonian can be written as
\begin{equation}\label{B2}
    2 m H= {p_1}^2 + \left(p_2- e B x_1 \right)^2
    + 2 m v_F \sigma_1 p_1 + 2 m v_F \sigma_2 \left(p_2 - e B x_1 \right)\,,
\end{equation}
which clearly commutes with $p_2$. This allows to look for generalized eigenfunctions of the form
\begin{equation}\label{B3}
    \psi(\mathbf{x})= \frac{e^{\imath k x_2}}{\sqrt{2\pi}} \,\Phi(x_1)\,,
\end{equation}
where $\Phi=\left(
              \begin{array}{c}
                \varphi \\
                \chi \\
              \end{array}
            \right)$.

The eigenvalue equation for the Hamiltonian, $(H-\mathcal{E})\psi=0$, reduces to the pair of coupled differential equations
\begin{equation}\label{B4}
    \begin{array}{c}\displaystyle
    \left\{\left[ {p_1}^2 + (e B)^2 \left( x_1 - \frac{k}{e B} \right)^2 \right] -\lambda \right\}\varphi=
    -2 m v_F \left\{ p_1 +\imath e B \left( x_1 - \frac{k}{e B}\right)\right\}\chi\,,
       \\ \\ \displaystyle
       \left\{\left[ {p_1}^2 + (e B)^2 \left( x_1 - \frac{k}{e B} \right)^2 \right] -\lambda \right\}\chi=
    -2 m v_F \left\{ p_1 -\imath e B \left( x_1 - \frac{k}{e B}\right)\right\}\varphi\,,
    \end{array}
\end{equation}
with $\lambda=2 m \mathcal{E}$.

At this point, it is convenient to change the remaining variable $x_1$ in favor of $q:= \sqrt{|e B|} \left(x_1-\frac{k}{e B} \right)$. So, $p_1=\sqrt{|e B|} p$, with $p:=-\imath \frac{\partial}{\partial q}$, and the system in Eq.\ (\ref{B4}) writes as
\begin{equation}\label{B5}
    \begin{array}{c}\displaystyle
    \left\{|e B|\left[ {p}^2 + q^2 \right] -\lambda \right\}\varphi(q)=
    -2 m v_F \sqrt{|e B|} \left\{ p +\imath \sign(e B)\, q\right\}\chi(q)\,,
       \\ \\ \displaystyle
       \left\{|e B|\left[ {p}^2 + q^2 \right] -\lambda \right\}\chi(q)=
    -2 m v_F \sqrt{|e B|} \left\{ p -\imath \sign(e B)\, q\right\}\varphi(q)\,.
    \end{array}
\end{equation}

\medskip

For simplicity, in the following we will take $e B>0$. The case $e B < 0$ can be obtained from the previous one by simply interchanging the components of the solutions, $\varphi(q) \leftrightarrow \chi(q)$, as can be easily seen from Eq.\ (\ref{B5}). So, we consider
\begin{equation}\label{BBB5}
    \left\{e B\left[ {p}^2 + q^2 \right] -\lambda \right\}\left(
                                                            \begin{array}{c}
                                                              \varphi(q) \\
                                                              \chi(q) \\
                                                            \end{array}
                                                          \right)
    =-2 m v_F \sqrt{e B} \left\{ p\, \sigma_1 -  q\, \sigma_2 \right\}\left(
                                                            \begin{array}{c}
                                                              \varphi(q) \\
                                                              \chi(q) \\
                                                            \end{array}
                                                          \right)\,.
\end{equation}

The presence of twice the Hamiltonian of a harmonic oscillator of frequency 1 in the brackets on the left hand side of Eqs.\ (\ref{BBB5}), operator with eigenfunctions $\phi_n(q)=e^{-q^2/2} H_n(q)$ ($H_n$ the Hermite polynomials
) and eigenvalues $2n+1$, for $n=0,1,\dots$, suggests that $\varphi(q)\sim \phi_n(q)$ and $\chi(q)\sim \phi_{n'}(q)$, for some $n,n'$. Moreover, since the Hermite functions satisfy
\begin{equation}\label{B6}
    \left\{\begin{array}{c} \displaystyle
             {\phi_n}'(q)+q \phi_n(q)= 2n \phi_{n-1}(q)
             \\ \\ \displaystyle
             {\phi_n}'(q)-q \phi_n(q)= - \phi_{n+1}(q)
           \end{array}
     \right.
\end{equation}
one concludes that these solutions are of the form
\begin{equation}\label{B7}
    \varphi(q)= c_1 \phi_{n+1}(y)\,, \qquad
    \chi(q)= c_2 \phi_{n}(q)\,,
\end{equation}
for $n=0,1,2,\cdots$, with $c_1,c_2\in \mathbb{C}$. Replaced in Eq.\ (\ref{B5}), we get the homogeneous system of algebraic equations
\begin{equation}\label{B8}
    M \left(
        \begin{array}{c}
          c_1 \\
          c_2 \\
        \end{array}
      \right)=\left(
                \begin{array}{c}
                  0 \\
                  0 \\
                \end{array}
              \right)\,,
\end{equation}
with
\begin{equation}\label{B9}
    M=\left(
          \begin{array}{cc}
            e B (2n+3)-\lambda &  \imath 2 m v_F\sqrt{e B} \\ \\
             - \imath 2 m v_F \sqrt{e B} (2n+2) & e B (2n+1)-\lambda \\ \\
          \end{array}
        \right)\,.
\end{equation}

The eigenvalues are determined by the condition
\begin{equation}\label{B10}
    {\rm det} M = \left[2 e B (n+1)-\lambda \right]^2 - (e B)^2
    -8 m^2 {v_F}^2 e B (n+1) =0\,,
\end{equation}
which implies that
\begin{equation}\label{B11}
    \begin{array}{c}\displaystyle
      \mathcal{E}_{n,s}=\frac{\lambda_{n,s}}{2 m}
      =\left(v_F \sqrt{e B}\right)  \frac{1}{z} \left[n+1+ \frac{s }{2} \sqrt{1+8 z^2 (n+1)} \right]\,,
    \end{array}
\end{equation}
with $s=\pm 1$ and
\begin{equation}\label{z}
    z:= \frac{m v_F}{\sqrt{e B}}\,.
\end{equation}


Appropriately normalized, the generalized eigenfunctions are written as
\begin{equation}\label{B14}
    \psi_{k,n,s}(\mathbf{x})= \frac{e^{\imath k x_2}}{\sqrt{2\pi}}\,
      K_{n,s} \, e^{-\frac{q^2}{2}} \left( \begin{array}{c}
       H_{n+1}(q) \\ \\ \displaystyle
      \frac{\imath}{2z}\left[ 1-s  \sqrt{1+8 z^2 (n+1)} \right] H_n(q)\\
      \end{array}
      \right)\,,
\end{equation}
where
\begin{equation}\label{B15}
    K_{n,s}=\sqrt[4]{e B}\ \frac{2^{-\frac{n}{2}-1} \sqrt{\left(\frac{s
   }{2}+\sqrt{1+8 z^2 (n+1)}\right)^2-\frac{1}{4}}}{\sqrt[4]{\pi }
   \sqrt{(n+1)!} \sqrt{1+8 z^2 (n+1)}}
\end{equation}
and
\begin{equation}\label{B16}
    q= \sqrt{e B} \left(x_1-\frac{k}{e B} \right)\,.
\end{equation}

Indeed, taking into account the orthogonality relations for the Hermite functions
it can be easily verified that
\begin{equation}\label{BB17}
    \left(\psi_{k,n,s},\psi_{k',n',s'}\right)=\delta_{n,n'} \delta_{s,s'}\delta(k-k')\,.
\end{equation}

\medskip

Finally, notice that there is another solution $\Phi_0$ of Eq.\ (\ref{BBB5}) whose components are given by
\begin{equation}\label{BB7}
    \varphi(q)= \left( \frac{e B}{\pi} \right)^{1/4} \phi_{0}(q)\,, \qquad
    \chi(q)= 0\,.
\end{equation}
Indeed, since $H_0(q)=1$, we have
\begin{equation}\label{BBB7}
    (p-\imath q) \phi_0(q)=-\imath (\partial_q+ q) e^{-q^2/2}
    =0
\end{equation}
and Eq.\ (\ref{B5}) reduces to
\begin{equation}\label{cero-1}
    \left\{\frac{e B}{m}\, \frac{1}{2}\left[ {p}^2 + q^2 \right] -\mathcal{E}_0 \right\}\phi_0(q)=
    \left\{\frac{e B}{m}\, \frac{1}{2} -\mathcal{E}_0 \right\}\phi_0(q)= 0\,.
\end{equation}
This implies that $\mathcal{E}_0=eB/2m$, which is independent of $v_F$.

One can also verify that
\begin{equation}\label{psi0}
    \psi_{k,0}(\mathbf{x})=\frac{e^{\imath k x_2}}{\sqrt{2\pi}}\left( \frac{e B}{\pi} \right)^{1/4}
    \left(
    \begin{array}{c}
    \phi_0(q) \\
    0 \\
    \end{array}
    \right)
\end{equation}
satisfy
\begin{equation}\label{BBB17}
    \left(\psi_{k,n,s},\psi_{k',0}\right)=0\,, \qquad \left(\psi_{k,0},\psi_{k',0}\right)=\delta(k-k')\,.
\end{equation}

\smallskip

As mentioned in Section \ref{NC-space}, these results can be interpreted as the solution of a non-Abelian version of the Landau problem, in which we have also introduced a constant and uniform non-Abelian magnetic field. Moreover, in the $m\rightarrow\infty$ limit, these eigenfunctions are similar to the solutions found \cite{Mariel} for the Dirac equation in  a constant magnetic background.

Notice that one can take appropriate linear combinations of the generalized eigenfunctions in Eq.\ (\ref{B14}) and (\ref{psi0}) so as to construct a manifestly complete set of generalized vectors in our space,
\begin{equation}\label{complete}
    \left\{ \frac{e^{\imath k x_2}}{\sqrt{2\pi}} \, \phi_n(q) \left(
                                                                \begin{array}{c}
                                                                  1 \\
                                                                  0 \\
                                                                \end{array}
                                                              \right)
    , \frac{e^{\imath k x_2}}{\sqrt{2\pi}} \, \phi_n(q) \left(
                                                                \begin{array}{c}
                                                                  0 \\
                                                                  1 \\
                                                                \end{array}
                                                              \right)
    ; k\in \mathbb{R}, n=0,1,2,\dots \right\}\,,
\end{equation}
where $\phi_n(q)$ are the Hermite functions. Then, the set of generalized eigenvectors of $H$ we found is also complete, which ensures that in our analysis we got the whole spectrum of the Hamiltonian.

\medskip

In the following, we consider the large-$z$ (large-$m$) limit we are interested in\footnote{For $z\ll 1$ we have equally spaced levels in each branch, with slightly different slopes,
\begin{equation}\label{B19}
       \mathcal{E}_{n,s}=\frac{ e B }{m}\left\{ \left(n +1+\frac{s}{2}\right)
        + 2s z^2 (n+1)+  O\left( z^4 \right) \right\}\,.
\end{equation}
For the eigenfunctions we get
\begin{equation}\label{B20}
    \begin{array}{c}\displaystyle
      \psi_{k,n,+1}(\mathbf{x})= \frac{e^{\imath k x_2}}{\sqrt{2\pi}}\,
      \frac{2^{-\frac{n}{2}-\frac{1}{2}}}{\sqrt[4]{\pi }
   \sqrt{(n+1)!}}
      \,e^{-\frac{q^2}{2}}\times
    \\ \\ \displaystyle
    \left(
\begin{array}{c}
\left[  1 -  z^2 (n+1) +O\left(z^4\right) \right] H_{n+1}(q)
 \\ \\ \displaystyle
 \left[-2 \imath  z (n+1) +6 \imath z^3  (n+1)^2
 +O\left(z^4\right) \right] H_n(q)
\end{array}
\right)
    \end{array}
\end{equation}
and
\begin{equation}\label{B20p}
    \begin{array}{c}\displaystyle
      \psi_{k,n,-1}(\mathbf{x})= \frac{e^{\imath k x_2}}{\sqrt{2\pi}}\,
      \frac{2^{-n/2} }{\sqrt[4]{\pi}\sqrt{n!}}
      \,e^{-\frac{q^2}{2}}\times
    \\ \\ \displaystyle
    \left(
\begin{array}{c}
 \left[z-3 z^3 (n+1) +O\left(z^5\right) \right] H_{n+1}(q)
 \\ \\ \displaystyle
 \imath \left[1 -   z^2  (n+1) +\frac{11}{2}\,
    z^4 (n+1)^2 +O\left(z^5\right)\right]H_n(q)
\end{array}
\right)\, .
\end{array}
\end{equation}

The $z\rightarrow 0$-limit is consistent with the usual Landau levels problem.}.

\medskip


For $z\gg 1$ the energies are given by
\begin{equation}\label{B17}
    \mathcal{E}_{n,s}= v_F\sqrt{{e B}}\left\{ s \sqrt{2(n+1)}
   +\frac{(n+1)}{z}+ O\left(z^{-2}\right) \right\}
\end{equation}
and the eigenfunctions reduce to
\begin{equation}\label{B18}
    \psi_{k,n,s}(\mathbf{x})= \frac{e^{\imath k x_2}}{\sqrt{2\pi}}\,
    \frac{2^{-\frac{n}{2}-1}\,e^{-\frac{q^2}{2}}}{\sqrt[4]{\pi } \sqrt{(n+1)!}}
    \left(
\begin{array}{c}
 H_{n+1}(q)+O\left(z^{-1}\right) \\ \\ \displaystyle
 -\frac{\imath s}{2}  \sqrt{8 (n+1)} \,
   H_n(q)+O\left(z^{-1}\right)
\end{array}
\right)\,.
\end{equation}

\subsection*{Negative mass}\label{negative-mass}

Since we are interested in the description of low energy states and the construction of the solutions of Eq.\ (\ref{B4}) is independent of the sign of $m$, we can explore the behavior of the system for $m<0$. Indeed, it is safe to change the sign of $m$ in the solutions, which corresponds to the replacement $z\rightarrow-w$, with $w=|m|v_F/\sqrt{e B}>0$, in the last line in  Eq.\ (\ref{B11}) maintaining $v_F>0$. We get (See Figure 2)
\begin{equation}\label{g10}
    \begin{array}{c}\displaystyle
      \mathcal{E}_{n,s}
      =-\left(v_F \sqrt{e B}\right)  \frac{1}{w} \left[n+1+ \frac{s }{2} \sqrt{1+8 w^2 (n+1)} \right]=
      \\ \\ \displaystyle
      = \left(v_F \sqrt{e B}\right) \left\{ -s \sqrt{2(n+1)}-\frac{n+1}{w}+ O\left(w^{-2}\right) \right\}
      \,,
    \end{array}
\end{equation}
\begin{figure} \label{m-negativo.fig}
\epsffile{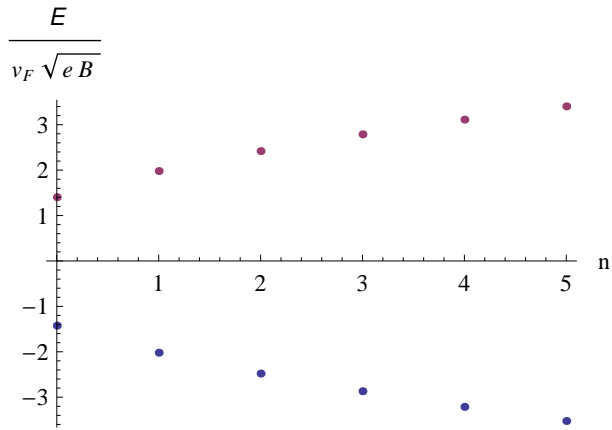}
\caption{The energy levels $\mathcal{E}_{n,s}$ for negative mass, in units of $v_F \sqrt{e B}$ and for $w=100$.}
\end{figure}

On the other hand\footnote{Notice that $\mathcal{E}_0\rightarrow 0$ for $m\rightarrow \infty$. This is consistent with the well known existence of zero modes in the spectrum of massless Dirac fermions coupled to gauge fields, discussed in different contexts in \cite{McClure}, \cite {AS} and \cite{rede}, for example.}, $\mathcal{E}_0=-eB/2|m|<0$.

\section{The relation with graphene}\label{relation-graphene}

The structure of graphene \cite{Castro} can be seen as a triangular lattice with a basis of two atoms per unit cell. The lattice vectors are
\begin{equation}\label{g1}
    \mathbf{a}_1=\frac{3}{2}\, a\left(
                              \begin{array}{c}
                                1 \\
                                \frac{1}{\sqrt{3}} \\
                              \end{array}
                            \right)
     \,, \quad \mathbf{a}_2=\frac{3}{2}\, a \left(
                              \begin{array}{c}
                                1 \\
                                \frac{-1}{\sqrt{3}} \\
                              \end{array}
                            \right)\,,
\end{equation}
where $a\approx 1.42 {\AA}$ is the lattice constant (distance between nearest neighbor carbon atoms). The vectors characterizing the reciprocal lattice (whose elementary cell is the Brillouin zone) are given by
\begin{equation}\label{g2}
    \mathbf{b}_1=\frac{2\pi}{3 a}\left(
                              \begin{array}{c}
                                1 \\
                                {\sqrt{3}} \\
                              \end{array}
                            \right)
     \,, \quad \mathbf{b}_2=\frac{2\pi}{3 a} \left(
                              \begin{array}{c}
                                1 \\
                                -{\sqrt{3}} \\
                              \end{array}
                            \right)\,.
\end{equation}

The superposed triangular lattices form an hexagonal honeycomb array of carbon atoms where the nearest-neighbor vectors are
\begin{equation}\label{g3}
    \boldsymbol{\delta}_1=\frac{a}{2}\left(
                              \begin{array}{c}
                                1 \\
                                {\sqrt{3}} \\
                              \end{array}
                            \right)\,, \quad
                                \boldsymbol{\delta}_2=\frac{a}{2}\left(
                              \begin{array}{c}
                                1 \\
                                -{\sqrt{3}} \\
                              \end{array}
                            \right)\,, \quad
                                \boldsymbol{\delta}_3={a}\left(
                              \begin{array}{c}
                                -1 \\
                                0 \\
                              \end{array}
                            \right)\,.
\end{equation}

The tight-binding model for graphene, where it is assumed that electrons can only hop to both nearest ($\langle i,j \rangle$) and next-to-nearest  ($\langle\langle i,j \rangle\rangle$) neighbor atoms, is described by the Hamiltonian \cite{Castro}
\begin{equation}\label{g4}
    \begin{array}{c} \displaystyle
      H=-t \sum_{\langle i,j \rangle} \sum_{s=\pm} \left(a^\dagger_{i,s} \, b_{j,s} + {\rm h.c.}\right)-
      \\ \\ \displaystyle
      - t' \sum_{\langle\langle i,j \rangle\rangle} \sum_{s=\pm}
      \left(a^\dagger_{i,s} \, a_{j,s} +b^\dagger_{i,s} \, b_{j,s} + {\rm h.c.}\right)\,,
    \end{array}
\end{equation}
where $t$ and $t'$ are the hoping energies, $a^\dagger_{i,s}$ and  $a_{i,s}$ are the creation and annihilation operators of electrons in the site $i$ with spin $s$ belonging to the triangular sublattice $A$, and similarly for the sublattice $B$.  By Fourier transforming these operators and diagonalizing the resulting expression, one gets the band structure (dispersion relations) \cite{Castro}
\begin{equation}\label{g5}
    E_{\pm}(\mathbf{k})= \pm t \sqrt{f(\mathbf{k})}- t' \left[f(\mathbf{k})-3\right]\,,
\end{equation}
with
\begin{equation}\label{g6}
    f(\mathbf{k})=3+4 \cos \left(\frac{3  k_1 a}{2}\right) \cos
   \left(\frac{\sqrt{3}}{2} k_2 a\right)+2 \cos
   \left(\sqrt{3} k_2 a\right) \geq 0\,.
\end{equation}

The minima of $f(\mathbf{k})$ in the Brillouin zone
correspond to the \emph{Dirac points}
\begin{equation}\label{g8}
    \mathbf{K}=\frac{2\pi}{3 a}\left(
                                 \begin{array}{c}
                                   1 \\
                                   \frac{1}{\sqrt{3}} \\
                                 \end{array}
                               \right)\,,
                               \quad
    \mathbf{K}'=\frac{2\pi}{3 a}\left(
                                 \begin{array}{c}
                                   1 \\
                                   \frac{-1}{\sqrt{3}} \\
                                 \end{array}
                               \right)\,,
\end{equation}
where $f(\mathbf{K})=0=f(\mathbf{K}')$.  Then, the bands corresponding to each sign in the dispersion relations, Ec.\ (\ref{g5}), touch each other at the Dirac points. Notice that these two independent Dirac points have been chosen so that they are related by the reflection of $\mathbf{k}$ about the $k_1$-axis ($k_2\rightarrow -k_2$) \cite{Castro}.

A series expansion of $E_\pm(\mathbf{k})$ around $\mathbf{k}=\mathbf{K}$ leads to
\begin{equation}\label{g9}
    E_s(\mathbf{K}+ \mathbf{k})=s\, t \left[ \frac{3 }{2}\, a |\mathbf{k}| -\frac{3}{8}\, a^2 \mathbf{k}^2 \sin (3 \theta ) \right]+ t' \left[-\frac{9}{4} \,  a^2 \mathbf{k}^2 +3\right]+O\left(|\mathbf{k}|^3\right)\,,
\end{equation}
where $\tan(\theta)=k_2/k_1$ and $s=\pm 1$.

\medskip

The expansion around the second Fermi point, $\mathbf{K}'$, leads to the same expression with $\mathbf{k}$ reflected about the $k_1$-axis. This corresponds to the change $\theta\rightarrow -\theta$, which changes the sign of the the second term in the first bracket in the right hand side of Eq.\ (\ref{g9}).

\medskip

We now turn to the comparison with the model developed in the previous Sections. From Eqs.\ (\ref{g9}) and (\ref{disp}), it is seen that we can identify  $v_F\equiv \frac{3}{2}\, a t$. Moreover, the next-to-nearest neighbor contribution can be represented in our (free) model by the \emph{mass} term through the identification $-\frac{9}{4}\, t' a^2 \equiv \frac{1}{2m}$, up to a rigid displacement of the spectrum in an energy $3 t'$ (which can be subtracted to restore the particle-hole symmetry at linear order in $|\mathbf{k}|$). This means that we must consider a \emph{negative mass} in our model, $m=- 2/\left(9 t'a^2\right) <0$\footnote{According to the values of these parameters reported in \cite{Castro}, we have \label{foot-w}
\begin{equation}\label{parameters}
    a=1.42 {\AA}\,, \quad t=2.8 eV\,, \quad t'=0.1 eV\,.
\end{equation}
Then,
\begin{equation}\label{valores}
    v_F=600\, {\rm Km/sec}\,,
\quad m c^2=-4.3\times 10^{6}\, {\rm eV}\,,
\end{equation}
or, in natural units ($v_F/c \rightarrow v_F$, $mc/\hbar\rightarrow m $),
\begin{equation}\label{valores-naturales}
    {v_F}=2 \times 10^{-3} \,, \quad m=-8.53 \times 10^{13} \, {\rm meter}^{-1}\,.
\end{equation}
This is a rather large value for the (negative) mass of the quasiparticles, since $|m|\simeq 32.9 \, m_e$. For example, for an external magnetic field $B=10$\, Tesla,  we get $w=\frac{|m| c}{\hbar} \, \frac{v_F}{c}\, \sqrt{\frac{\hbar}{e B}}=1385.$}.

\medskip

On the other hand, the second (quadratic) term in the nearest neighbor contribution depends on the direction of $\mathbf{k}$. We can not reproduce this behavior within the framework of our model since it is rotationally invariant. Rather, we must treat it as a perturbation on the solutions we found for our model. This will be considered later.

\medskip

Then, we will assume that the effective Hamiltonian for the low energy states of this system around $\mathbf{K}$ in the presence of an electromagnetic field, $H_{\mathbf{K}}$, is obtained though minimal coupling and coincides with the one given in Eq.\ (\ref{Hma-1}).

Since the application of our model to graphene requires $w \approx 10^3$ (See footnote \ref{foot-w}), the large $w$ expansion in Eq.\ (\ref{g10}) gives a good description of the self-energies. So, we get for quasi-particle states
\begin{equation}\label{gg10}
      \mathcal{E}_{n,-}
      = \left(v_F \sqrt{e B}\right) \left\{ \sqrt{2(n+1)}-\frac{n+1}{w}+ O\left(w^{-2}\right) \right\}
      \,,
\end{equation}
and for hole states
\begin{equation}\label{ggg10}
      \mathcal{E}_{n,+}
      = \left(v_F \sqrt{e B}\right) \left\{ - \sqrt{2(n+1)}-\frac{n+1}{w}+ O\left(w^{-2}\right) \right\}
      \,.
\end{equation}
Moreover, there is an additional (hole) state at energy $\mathcal{E}_0=- \frac{v_F\sqrt{e B}}{2 w}<0$, slightly below the zero energy level (see Eq.\ (\ref{cero-1})). Indeed, the difference in energy for the two first hole states is $v_F \sqrt{e B}\left( \sqrt{2}+O(w^{-1}) \right)$\footnote{For $w=10^{3}$ and $B=10$\, Tesla, we have in full units
\begin{equation}\label{two-holes}
    v_F\sqrt{e B}\rightarrow \hbar c \, \frac{v_F}{c} \sqrt{\frac{e B}{\hbar}} = 4.87 \times {10}^{-2}\, {\rm eV}\,.
\end{equation}
This leads to $\mathcal{E}_0=-2.43 \times {10}^{-5}\, {\rm eV}$ and $\mathcal{E}_{0,+}=-6.89\times {10}^{-2}\, {\rm eV}$.}.

\subsection{The second Fermi point contribution}\label{AIQHE}

As previously mentioned, the dispersion relations near the Fermi points are related by the reflection of momenta about the $k_1$-axis. Then, for the linear terms in the effective free Hamiltonian describing the states around the second Fermi point we can write \cite{Castro}
\begin{equation}\label{second-1}
    v_F \left( \sigma_1 p_1 - \sigma_2 p_2 \right) = v_F\, \BS^* \cdot \mathbf{p}
    = - \sigma_2 \left(v_F\, \BS \cdot \mathbf{p} \right) \sigma_2\,.
\end{equation}

Accordingly, we will assume that the effective Hamiltonian describing the quasi-particles near the second Fermi point in our model, $H_{\mathbf{K}'}$, can be obtained from $H_{\mathbf{K}}$ through the transformation
\begin{equation}\label{second-2}
    H_{\mathbf{K}'}=- \sigma_2 H_{\mathbf{K}} \sigma_2=
    - \frac{\left( \mathbf{p}-e \mathbf{A} \right)^{2}}{2 m}
    + v_F \, \BS^* \cdot \left( \mathbf{p}-e \mathbf{A} \right)\,.
\end{equation}

Notice that both $H_{\mathbf{K}}$ and $H_{\mathbf{K}'}$ are left invariants under time-reversal times parity transformations, $\mathcal{TP}$ (See the discussion in Ref.\ \cite{Vozmediano} about the symmetries of this system). Indeed \cite{Vozmediano},
\begin{equation}\label{TP1}
    \begin{array}{c}\displaystyle
      \left(\mathcal{TP} \right) H_{\mathbf{K}} \left( \mathcal{TP} \right)^\dagger
    =\sigma_1 {H_{\mathbf{K}}}^* \sigma_1=
    \\ \\ \displaystyle
      = \sigma_1 \left\{
    \frac{\left( \mathbf{p}-e \mathbf{A} \right)^{2}}{2 m}
    + v_F \, \BS^* \cdot \left( \mathbf{p}-e \mathbf{A} \right)\right\} \sigma_1=H_{\mathbf{K}}
    \end{array}
\end{equation}
and similarly
\begin{equation}\label{TP2}
    \begin{array}{c} \displaystyle
          \left(\mathcal{TP} \right) H_{\mathbf{K}'} \left( \mathcal{TP} \right)^\dagger
    =-{\sigma_2}^*\left(\mathcal{TP} \right) H_{\mathbf{K}} \left( \mathcal{TP} \right)^\dagger
    {\sigma_2}^*=
    \\ \\ \displaystyle
      =- \sigma_2 H_{\mathbf{K}} \sigma_2= H_{\mathbf{K}'}\,.
    \end{array}
\end{equation}

\smallskip

Evidently, the spectrum of $H_{\mathbf{K}'}$ is obtained from that of $H_{\mathbf{K}}$ by a reflection about the origin and its (generalized) eigenfunction are just $\sigma_2$ times those of $H_{\mathbf{K}}$. This interchanges quasi-particles with holes. Indeed,
\begin{equation}\label{second-3}
   H_{\mathbf{K}'} \left(\sigma_2 \psi_{k,n,s}\right) = -\sigma_2 \left(H_{\mathbf{K}} \psi_{k,n,s}\right)
    =- \mathcal{E}_{n,s} \left(\sigma_2 \psi_{k,n,s}\right)\,.
\end{equation}
Moreover,
\begin{equation}\label{second-4}
    H_{\mathbf{K}'}  \left(\sigma_2 \psi_{0}\right)=-\sigma_2 \left(H_{\mathbf{K}} \psi_{0}\right)
    =- \mathcal{E}_{0} \left(\sigma_2 \psi_{0}\right)\,,
\end{equation}
with an energy $\mathcal{E'}_{0} =- \mathcal{E}_{0} = \frac{v_F\sqrt{e B}}{2 w}>0$. This state corresponds to a quasi-particle of energy slightly above zero.

\medskip

Therefore, taking into account the eigenstates of both $H_{\mathbf{K}}$ and $H_{\mathbf{K}'}$ we get, for quasi-particles and for holes, an almost doubly degenerate spectrum, except for one state of quasi-particle and one state of hole with energies near zero. For positive energy states we have
\begin{equation}\label{second-5}
    \begin{array}{c}\displaystyle
      \mathcal{E}_{n,-}
      = \left(v_F \sqrt{e B}\right) \left\{ \sqrt{2(n+1)}-\frac{n+1}{w}+ O\left(w^{-2}\right) \right\}
      \,,
      \\ \\ \displaystyle
      -\mathcal{E}_{n,+}
      = \left(v_F \sqrt{e B}\right) \left\{  \sqrt{2(n+1)} + \frac{n+1}{w}+ O\left(w^{-2}\right) \right\}
      \,,
    \end{array}
\end{equation}
where the gap between contiguous states is
\begin{equation}\label{second-6}
    \triangle \mathcal{E}_n = 2 \left(v_F \sqrt{e B}\right)
    \left\{\frac{n+1}{w}+ O\left(w^{-2}\right) \right\}\,.
\end{equation}

As we discuss in the next Section, this spectrum reproduces qualitatively the anomalous integer quantum Hall effect found in graphene \cite{Gusynin-2005,Peres-2006,Novo-2005,Zhang}, which shows a nonvanishing Hall conductivity for a small (positive or negative) Fermi level.

\section{The Hall conductivity}\label{Hall-conductivity}

As discussed in Appendix \ref{Ec-Mov}, the Hall conductivity has a topological character and can be calculated from the weak field and gradient expansion of the effective action of the system. However, since we know the exact eigenvalues of energy of our model, we will employ a more direct evaluation method, based on the relation established between the conserved current and the external electromagnetic field, Ecs.\ (\ref{L14}) and (\ref{L15}).

For our purposes, it will be sufficient to consider the mean value of the density $j_0(x)$. Then, as in the previous Sections, we will take $\mathbf{E}=0$ and $B\neq 0$ perpendicular to the plane of the system.

Let us start by writing down the partition function of the particles around one Fermi point at inverse temperature $\beta$ and chemical potential $\mu$. It can be obtained through a Wick rotation of the generating functional $Z[A]$ (see Eq.\ (\ref{L9})) to the  2+1 Euclidean space by means of the replacements $x^0=t\rightarrow-\imath \tau$, $A_0\rightarrow \imath A_3$, maintaining unaltered the other coordinates and components of the gauge field.

Since, in our case, $A_0=0=A_1$, $A_2=B x_1$, from Eqs.\ (\ref{L1}) and (\ref{L9}), this Wick rotation leads to the partition function
\begin{equation}\label{pf1}
    \mathcal{Z}(\beta,\mu,B):=   \int \mathcal{D} \psi^\dagger \mathcal{D}\psi \, e^{\displaystyle
   \int_0^\beta dt \int d^2 x \, \psi^\dagger\left\{-\frac{\partial}{\partial \tau}+ \mu -H\right\}\psi}\,,
\end{equation}
where $H$ is $H_{\mathbf{K}}$ (or $H_{\mathbf{K}'}$) and the functional integral is performed on the set of configurations of the fermionic field which satisfy anti-periodic boundary conditions on $[0,\beta]$. Then,
\begin{equation}\label{pf2}
    \begin{array}{c} \displaystyle
      \frac{\partial \log \mathcal{Z}}{\partial \mu}(\beta,\mu,B)=\int_0^\beta dt \int d^2 x \left\langle \psi^\dagger
    \psi \right\rangle =
    \\ \\ \displaystyle
      = \beta \int d^2 x \ \frac{1}{e} \, J_0(\beta,\mu,B)  = \beta \int d^2 x \ \frac{\sigma_{xy}}{e} \, B\,,
    \end{array}
\end{equation}
where use has been made of Eq.\ (\ref{L15}).

\smallskip

Our goal is now to evaluate the partition function as the functional determinant
\begin{equation}\label{pf3}
    \mathcal{Z}(\beta,\mu,B) = {\rm Det}\left(D\right)
\end{equation}
where $D=-\frac{\partial}{\partial \tau}+ \mu -H$ is a differential operator defined on a domain of anti-periodic functions of $\tau\in [0,\beta]$. Even though $D$ is not symmetric, since $H$ does not depend on $\tau$ this operator has a complete set of orthogonal generalized eigenfunctions constructed as
\begin{equation}\label{pf4}
   \Psi_{l,k,n,s} = \frac{1}{\sqrt{\beta}}\, e^{\displaystyle-\imath \omega_l \tau}\, \psi_{k,n,s}\,,
   \quad \Psi_{l,k,0}=\frac{1}{\sqrt{\beta}}\, e^{\displaystyle-\imath \omega_l \tau}\, \psi_{k,0}\,,
\end{equation}
with $l\in \mathbb{Z}$, $k\in \mathbb{R}$, $n=0,1,2,\dots$, $s=\pm 1$, and where the Matsubara frequencies
\begin{equation}\label{pf5}
    \omega_l= \frac{2\pi}{\beta}\left(l+\frac{1}{2}\right)\,,\quad {\rm with}\quad\omega_l=-\omega_{-l-1}\,.
\end{equation}
Then, the eigenvalues of $D$ are given by
\begin{equation}\label{pf6}
    \begin{array}{c} \displaystyle
          D \Psi_{l,k,n,s}= \lambda_{l,n,s} \Psi_{l,k,n,s}\,, \qquad
    \lambda_{l,n,s}=\imath \omega_l +\mu - \mathcal{E}_{n,s}\,,
    \\ \\ \displaystyle
    D \Psi_{l,k,0}= \lambda_{l,0} \Psi_{l,k,0}\,, \qquad
    \lambda_{l,0}=\imath \omega_l +\mu - \mathcal{E}_{0}\,,
    \end{array}
\end{equation}
where $\mathcal{E}_{n,s}$ and $\mathcal{E}_{0}$ are the eigenvalues of $H$, studied in the previous Sections.

Notice that $\lambda_{l,n,s}$ and $\lambda_{l,0}$ are independent of $k$. Therefore, in evaluating ${\rm Det}(D)$ we will forget about the index $k$ and, at the end, take into account the degeneracy it introduces \emph{per unit area}, given by the number of flux quanta per unit area \cite{Fradkin,Wen}, $\Delta=eB/2\pi$ (or $\Delta=eB/h$ in full units).

We define ${\rm Det}(D)$ as the (gauge invariant) $\zeta$-function determinant \cite{Dow,Seeley},
\begin{equation}\label{pf7}
    \log{\rm Det}(D):= \left.- \frac{d}{du} {\rm Tr}\left\{\left(\frac{D}{\Lambda}\right)^{-u}\right\}\right|_{ u \rightarrow 0}\,,
\end{equation}
where $\Lambda$ is an arbitrary mass scale, the trace is evaluated for $\Re (u)$ sufficiently large and ${ u \rightarrow 0}$ stands for the analytic continuation of its derivative to a neighborhood of $u=0$. In our case,
\begin{equation}\label{pf8}
    \mathcal{Z}(\beta,\mu,B)=\left. - \frac{d}{du} \left\{\sum_{l,n,s} \left(\frac{\lambda_{l,n,s}}{\Lambda}\right)^{-u}
    + \sum_{l} \left(\frac{\lambda_{l,0}}{\Lambda}\right)^{-u}\right\}\right|_{ u \rightarrow 0}\,.
\end{equation}

Let us first consider the contribution of the second term on the right hand side and evaluate
\begin{equation}\label{pf9}
    \sum_{l=-\infty}^\infty \left(\frac{\lambda_{l,0}}{\Lambda}\right)^{-u}=\sum_{l=0}^\infty \left(\frac{\imath \omega_{l}+\mu-\mathcal{E}_0}{\Lambda}\right)^{-u}+\sum_{l=0}^\infty \left(\frac{-\imath \omega_{l}+\mu-\mathcal{E}_0}{\Lambda}\right)^{-u}\,,
\end{equation}
where use has been made of the properties of the Matsubara frequencies stated in Eq.\ (\ref{pf5}). We can also write
\begin{equation}\label{pf10}
    \begin{array}{c} \displaystyle
      \sum_{l=-\infty}^\infty \left(\frac{\lambda_{l,0}}{\Lambda}\right)^{-u}=\left(\frac{2\pi}{\beta \Lambda}\right)^{-u}
    \left\{\sum_{l=0}^\infty \left[{\imath \left(l+\frac{1}{2}\right)+\frac{\beta}{2\pi}\left(\mu-\mathcal{E}_0\right)}\right]^{-u}+\right.
    \\ \\ \displaystyle
      \left. +\sum_{l=0}^\infty \left[{-\imath \left(l+\frac{1}{2}\right) +\frac{\beta}{2\pi}\left(\mu-\mathcal{E}_0\right)}\right]^{-u}\right\}\,.
    \end{array}
\end{equation}

These series can be expressed in terms of the Hurwitz $\zeta$-function \cite{Mathematica,MathWorld}. Taking into account that this function has branch cut discontinuities in the complex  plane of its second argument running from 0 to $-\infty$, we get
\begin{equation}\label{pf11}
    \begin{array}{c} \displaystyle
      \sum_{l=-\infty}^\infty \left(\frac{\lambda_{l,0}}{\Lambda}\right)^{-u}=\left(\frac{2\pi}{\beta \Lambda}\right)^{-u}
    \left\{e^{\displaystyle-\imath \frac{\pi}{2} u} \zeta\left(u,\frac{1}{2}+e^{\displaystyle-\imath\frac{\pi}{2}\, {\rm sign} \left(\mu-\mathcal{E}_0\right)} \frac{\beta}{2\pi}\left|\mu-\mathcal{E}_0\right| \right)+\right.
    \\ \\ \displaystyle
      \left. +e^{\displaystyle\imath \frac{\pi}{2} u} \zeta\left(u,\frac{1}{2}+e^{\displaystyle\imath\frac{\pi}{2}\, {\rm sign} \left(\mu-\mathcal{E}_0\right)} \frac{\beta}{2\pi}\left|\mu-\mathcal{E}_0\right| \right) \right\}\,.
    \end{array}
\end{equation}

The expression in braces has a vanishing analytic continuation at $u=0$. As a consequence, the derivative of the right hand side at $u=0$ does not depend on the arbitrary scale $\Lambda$. Then, the contribution of this Landau level to $\log {\rm Det} D$ \emph{per unit area} and at low temperature is given by \cite{Mathematica,MathWorld}
\begin{equation}\label{pf12}
    \Delta \left\{ \left.- \frac{d}{du}\sum_{l=-\infty}^\infty \left(\frac{\lambda_{l,0}}{\Lambda}\right)^{-u} \right|_{ u \rightarrow 0} \right\}
    = \frac{eB}{2\pi}\left\{\beta \left( \mathcal{E}_0 - \mu \right)
    \Theta\left( \mathcal{E}_0 - \mu \right)
    + o\left( \frac{1}{\beta \left|\mu-\mathcal{E}_0\right|} \right)\right\}\,.
\end{equation}

Taking into account that we are interested in the mean number of particles with respect to the neutral material (\emph{i.e.}, with all the Landau levels with negative energy filled, which corresponds to $\mu=0$), from Eq.\ (\ref{pf2}), we finally get for the contribution of the Landau level with energy $\mathcal{E}_0$ to the Hall conductivity at zero temperature
\begin{equation}\label{pf13}
    \begin{array}{c} \displaystyle
      \left.   \frac{B}{e}\, \sigma_{xy} \right|_{\mathcal{E}_0}
   =   \frac{Be}{2\pi}\left\{ \frac{\partial}{\partial\mu}\left[ \left( \mathcal{E}_0 - \mu \right)
    \Theta\left( \mathcal{E}_0 - \mu \right) \right] -
    \left.\frac{\partial}{\partial\mu}\left[ \left( \mathcal{E}_0 - \mu \right)
    \Theta\left( \mathcal{E}_0 - \mu \right) \right]\right|_{\mu=0}
   \right\}
   \\ \\ \displaystyle
      = \frac{Be}{2\pi}\left\{ -
    \Theta\left( \mathcal{E}_0 - \mu \right) +
    \Theta\left( \mathcal{E}_0 \right)
   \right\}
   \\ \\ \displaystyle
      = \frac{Be}{2\pi}\big\{ \Theta(\mathcal{E}_0) \Theta\left(\mu- \mathcal{E}_0\right) -
      \Theta\left(  - \mathcal{E}_0 \right)     \Theta\left( \mathcal{E}_0 -\mu \right)
   \big\}
    \end{array}
\end{equation}
Therefore, for positive $\mathcal{E}_0$, we get a contribution $(+\frac{e^2}{2\pi})$ to $\sigma_{xy}$ if $\mu>\mathcal{E}_0$ and zero otherwise. On the other hand, for negative $\mathcal{E}_0$ we get a contribution $(-\frac{e^2}{2\pi})$  if $\mu<\mathcal{E}_0$ and zero otherwise.

At this point, it is clear that a similar result is obtained for any other Landau level. Indeed, the same considerations can be done for any other energy eigenvalue $\mathcal{E}$. Then, for a given chemical potential $\mu$, there is only a finite number of nonvanishing contributions and the Hall conductivity reduces to
\begin{equation}\label{pf14}
    \left. \sigma_{xy} \right|_{H_{\mathbf{K}}} = \frac{e^2}{2\pi}\left\{
    \Theta(\mu) \left(\sum_{0<\mathcal{E}<\mu} 1 \right)-
    \Theta(-\mu) \left(\sum_{\mu<\mathcal{E}<0} 1 \right)\right\}\,,
\end{equation}
where the sums run over the energy levels. Indeed, this is a rather general result for the Landau problem of a fermionic system.

\medskip

For the model of interest in the present paper, the spectrum to be considered is the union of those of $H_{\mathbf{K}}$ and $H_{\mathbf{K}'}$, $\mathcal{E}\in{\rm Spec} H_{\mathbf{K}} \bigcup {\rm Spec} H_{\mathbf{K}'}$, which takes into account the states around both Fermi points. But they must be taken with an additional degeneracy corresponding to the two polarizations of the electron spin, which has played no role up to now in this analysis.

In full units we get
\begin{equation}\label{pf15}
     \sigma_{xy} = \frac{2 e^2}{h}\left\{
    \Theta(\mu) \left(\sum_{0<\mathcal{E}<\mu} 1 \right)-
    \Theta(-\mu) \left(\sum_{\mu<\mathcal{E}<0} 1 \right)\right\}\,.
\end{equation}
Notice that
\begin{equation}\label{pf16}
    \left| \frac{h}{4 e^2} \, \sigma_{xy} \right| = \frac{1}{2}
\end{equation}
for small $|\mu|$ (but $|\mu|>|\mathcal{E}_0| = \frac{v_F\sqrt{e B}}{2 w}$). This is the characteristic of the \emph{anomalous integer quantum Hall effect} found in graphene \cite{Gusynin-2005,Novo-2005,Peres-2006,Zhang}, which shows a nonvanishing Hall conductivity for small (positive or negative) Fermi level.

\smallskip

In Figure 3, the Hall conductivity of this model as a function of $\mu$ (equal to the Fermi level $\epsilon_F$ at zero temperature) has been plotted for $w=10^{3}$. Notice that it is not possible to appreciate the structure of each step as a double step (due to the small degeneracy breaking remarked in Eqs.\ (\ref{second-5}) and (\ref{second-6})) with such a realistic value of $w$ (See Figure 4). Also notice that the Hall conductivity of our model vanishes in a small neighborhood of the origin, for $|\mu|< \frac{v_F\sqrt{e B}}{2 w}$.
\begin{figure} \label{sigmaXY.fig}
\epsffile{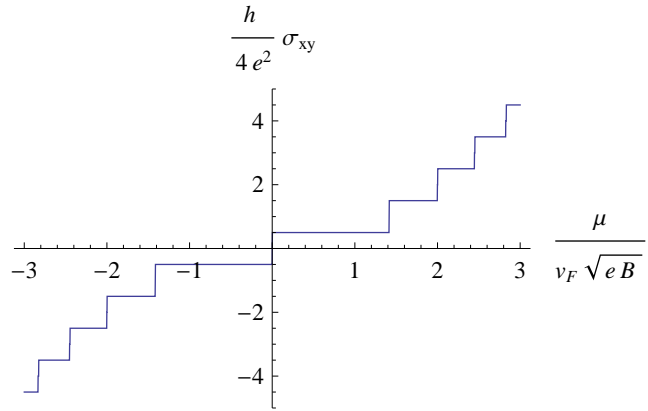}
\caption{The Hall conductivity for a realistic value of $w={10}^{3}$.}
\end{figure}
\begin{figure} \label{step.fig}
\epsffile{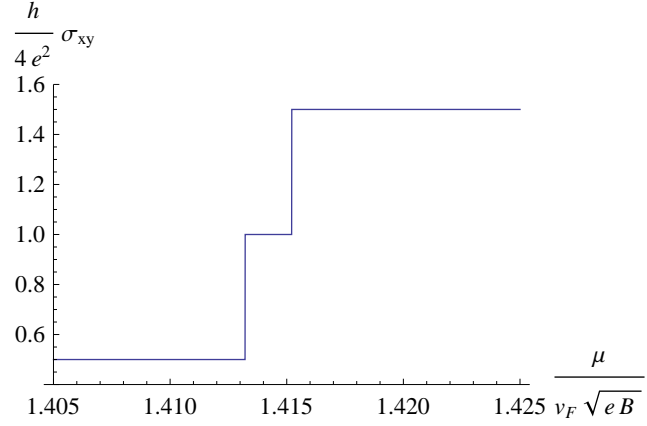}
\caption{The structure of the step in the Hall conductivity from $\frac{1}{2}$ to $\frac{3}{2}$, with $w={10}^{3}$.}
\end{figure}

\section{Perturbation with the quadratic nearest neighbor interaction term} \label{perturbation}

Notice that the quadratic term in the nearest neighbor dispersion relation in Eq.\ (\ref{g9}) can be obtained from the solutions of the free problem, Eqs.\ (\ref{eigen}) and (\ref{disp-2}), through
the (rotational symmetry breaking) perturbation
\begin{equation}\label{g11}
    \triangle H =- \frac{a}{4}\, v_F \, {p_2} \left[4 {p_1}^2 -\mathbf{p}^2\right] \left({\mathbf{p}}^2\right)^{-1}
    \left(\BS\cdot \mathbf{p}\right)\,.
\end{equation}
Indeed,
\begin{equation}\label{g12}
    \begin{array}{c} \displaystyle
      \triangle H \, \psi_{\mathbf{k}}(\mathbf{x})
    = - \frac{a}{4}\, v_F e^{\imath \mathbf{k}\cdot \mathbf{x}}\,{k_2} \left[4  {k_1}^2 -\mathbf{k}^2\right] \left( \mathbf{k}^2\right)^{-1/2} \left(\BS\cdot \mathbf{\hat{k}}\right) \chi_s(\mathbf{k})=
    \\ \\ \displaystyle
      =- s   \frac{a}{4}\, v_F\, {k_2} \left[4  {k_1}^2 -\mathbf{k}^2\right] \left( \mathbf{k}^2\right)^{-1/2} \, \psi_s(\mathbf{k})=
      -s \frac{3}{8}\, t\,a^2  \mathbf{k}^2 \sin\left(3 \theta \right)\, \psi_s(\mathbf{k})\,.
    \end{array}
\end{equation}

We will assume that, in the presence of an electromagnetic field $\mathbf{A}$, derivatives change into covariant derivatives and the perturbation just turns into the Hermitian expression
\begin{equation}\label{g13}
    \begin{array}{c} \displaystyle
      \triangle H =-  \frac{a}{8} v_F \left\{{\left(p_2-e A_2\right)}\left[4 {\left(p_1-e A_1\right)}^{{2}}
      -  \left(\mathbf{p}- e \mathbf{A}\right)^2 \right]\left[ \left(\mathbf{p}- e \mathbf{A}\right)^2\right]^{-1}\BS\cdot \left(\mathbf{p}-e \mathbf{A} \right)
      \right.
      \\ \\ \displaystyle
      \left.+\BS\cdot \left(\mathbf{p}-e \mathbf{A} \right)
      \left[ \left(\mathbf{p}- e \mathbf{A}\right)^2\right]^{-1}
      \left[4 {\left(p_1-e A_1\right)}^{{2}}
      -  \left(\mathbf{p}- e \mathbf{A}\right)^2 \right]
      {\left(p_2-e A_2\right)}
      \right\}\,.
    \end{array}
\end{equation}
Notice that there is an order indeterminacy in this definition. We will come back later on this point.

\medskip

We now consider a constant magnetic field as in Eq.\ (\ref{B1}). Then,
\begin{equation}\label{g14}
    p_1-e A_1=p_1=\sqrt{e B}\, p \,,\quad p_2-e A_2=p_2-e B x_1= -\sqrt{e B}\, q\,,
\end{equation}
and $\triangle H$ reduces to
\begin{equation}\label{g15}
    \begin{array}{c} \displaystyle
      \triangle H
= \mathfrak{H}+\mathfrak{H}^\dagger\,,
    \end{array}
\end{equation}
where
\begin{equation}\label{g16}
    \begin{array}{c} \displaystyle
      \mathfrak{H} = -\frac{a}{8} v_F\sqrt{e B} \, q \left[4 {p}^{{2}}
      - \left({p}^2+q^2\right) \right]
      \left[{p}^2+q^2 \right]^{-1}\left[p \sigma_1 - q \sigma_2\right]\,.
    \end{array}
\end{equation}

When $ \mathfrak{H}$ is applied to the functions in Eq.\ (\ref{B14}), taking into account that (See Eq.\ (\ref{BBB5}))
\begin{equation}\label{g17}
    \left[p \sigma_1 - q \sigma_2\right] \psi_{k,n,s}(\mathbf{x})
    = -\frac{1}{ w}\left\{\frac{m}{e B} \, \mathcal{E}_{n,s} - \frac{1}{2} \left( p^2 + q^2 \right)\right\}
     \psi_{k,n,s}(\mathbf{x})\,,
\end{equation}
where $\mathcal{E}_{n,s}$ is given in Eq.\ (\ref{g10}), we get
\begin{equation}\label{g18}
        \begin{array}{c} \displaystyle
       \mathfrak{H} \psi_{k,n,s}(\mathbf{x})
      = \frac{a}{8}\, v_F\sqrt{e B}\, \frac{K_{n,s}}{w}\,\frac{e^{\imath k x_2}}{\sqrt{2\pi}}
        \,  q \, e^{-\frac{q^2}{2}}\times
      \\ \\ \displaystyle
       \left( \begin{array}{c} \displaystyle
      \frac{\left\{\frac{m}{e B} \, \mathcal{E}_{n,s} - \frac{1}{2} \left( 2n+3 \right)\right\}}{\left[2n+3\right]} \left[4 q^{{2}} - 3 (2n+3)  \right] H_{n+1}(q)
      \\ \\ \displaystyle
      -\imath \left[\frac{ 1-s  \sqrt{1+8 w^2 (n+1)} }{2w}\right]
      \frac{\left\{\frac{m}{e B} \, \mathcal{E}_{n,s} - \frac{1}{2} \left( 2n+1 \right)\right\}}{\left[2n+1 \right]}
      \left[4 q^{{2}} - 3 (2n+1) \right] H_n(q)\\
      \end{array}
      \right)\,.
    \end{array}
\end{equation}

The first order perturbation on the eigenvalues of the Hamiltonian are given by the matrix elements
\begin{equation}\label{g19}
    \begin{array}{c}\displaystyle
      \left( \psi_{k',n,s},\triangle H \psi_{k,n,s}\right)=
    \left( \psi_{k',n,s}, \mathfrak{H} \psi_{k,n,s}\right)+\left( \mathfrak{H}\psi_{k',n,s},  \psi_{k,n,s}\right)
    \\ \\ \displaystyle
      = \delta(k-k')\, {a}\, v_F\sqrt{e B}\, \frac{{K_{n,s}}^2}{w}\,\times
      \\ \\ \displaystyle
      \int_{-\infty}^{\infty} dq\, e^{q^2} q \left\{
      \frac{\left\{\frac{m}{e B} \, \mathcal{E}_{n,s} - \frac{1}{2} \left( 2n+3 \right)\right\}}{\left[2n+3\right]} \left[4 q^{{2}} - 3 (2n+3)  \right] {H_{n+1}}^2(q) +
      \right.
      \\ \\ \displaystyle
     \left. + \left[\frac{ 1-s  \sqrt{1+8 w^2 (n+1)} }{2w}\right]^2
      \frac{\left\{\frac{m}{e B} \, \mathcal{E}_{n,s} - \frac{1}{2} \left( 2n+1 \right)\right\}}{\left[2n+1 \right]}
      \left[4 q^{{2}} - 3 (2n+1) \right] {H_n}^2(q) \right\}=0\,,
    \end{array}
\end{equation}
since it is the integral on the whole line of an odd function of $q$.

As previously noticed,  there is an order indeterminacy in the expression of $\mathfrak{H}$ in Eq.\ (\ref{g16}), since the operators in the right hand side of Eq.\ (\ref{g11}) commute. But any order one chooses will lead to an odd in $q$ integrand in the expression of $\left( \psi_{k',n,s},\triangle H \psi_{k,n,s}\right)$, giving then a vanishing result.

It is also immediate to show that $\mathfrak{H} \psi_0=0$, with $\psi_0$ given in Eq.\ (\ref{psi0}).

\smallskip

Therefore, we get no modification of the energy eigenvalues at first order in perturbation theory with $\triangle H$ in Eq.\ (\ref{g13}) (or with any permutation of the factors in $\mathfrak{H}$ in Eq.\ (\ref{g16})). This means that, in the presence of a constant magnetic field orthogonal to the plane, the spectrum of the Hamiltonian of our model seems to be rather stable against the rotational symmetry breaking perturbation in Eq.\ (\ref{g15}).

In particular, the Hall conductivity of the model described in Section \ref{Hall-conductivity} and the explanation of the anomalous integer quantum Hall effect based on it is not modified by the first order perturbation theory in $\triangle H$, operator which incorporates the (next to leading) quadratic term in the nearest neighbor interactions.

\section{Crossed Magnetic and electric fields}\label{crossed}

Let us consider also an electric field in the direction of $x_1$. So, we can take the electromagnetic field $A_\mu$ as
\begin{equation}\label{ME1}
    \mathbf{A}:= B x_1 \, \hat{\mathbf{e}}_2\,, \quad A_0:= - E x_1\,,
\end{equation}
also in the Coulomb gauge.

In this case the Hamiltonian is
\begin{equation}\label{ME2}
    H=\frac{\left( \mathbf{p}-e \mathbf{A} \right)^{2}}{2 m}
    + v_F \, \BS \cdot \left( \mathbf{p}-e \mathbf{A} \right) - e A_0\,.
\end{equation}
In the $m\rightarrow \infty$ limit a pseudo-Lorentz transformation of the linear Hamiltonian allows to solve the eigenvalue equation for crossed electric and magnetic fields in terms of the solutions of the case in which there is only a magnetic field applied perpendicularly to the plane of graphene \cite{Lukose,Castro-Peres,Mariel2}. But, in this nonrelativistic model, the quadratic mass term prevents to get the solution in this way.

\smallskip

Translation invariance in the $x_2$ direction suggests that the solutions are again of the form  $\Psi(\mathbf{x})=\frac{e^{\imath k x_2}}{\sqrt{2\pi}} \left( \begin{array}{c} \varphi(x_1) \\ \chi(x_1) \\ \end{array}\right)$, and the eigenvalue equation reduces to
\begin{equation}\label{ME3}
    \begin{array}{c}\displaystyle
    \left\{\left[ {p_1}^2 + (e B)^2 \left( x_1 - \frac{k}{e B} \right)^2 \right] -\lambda +2 m e E x_1\right\}\varphi=
    \\ \\ \displaystyle =
    -2 m v_F \left\{ p_1 +\imath e B \left( x_1 - \frac{k}{e B}\right)\right\}\chi\,,
       \\ \\ \displaystyle
       \left\{\left[ {p_1}^2 + (e B)^2 \left( x_1 - \frac{k}{e B} \right)^2 \right] -\lambda +2 m e E x_1\right\}\chi=
       \\ \\ \displaystyle =
    -2 m v_F \left\{ p_1 -\imath e B \left( x_1 - \frac{k}{e B}\right)\right\}\varphi\,,
    \end{array}
\end{equation}
with $\lambda=2 m \mathcal{E}$.

\smallskip

It is convenient to define a new variable as $q:= \sqrt{e B} \left[x_1-\frac{k}{e B}+\frac{mE}{e B^2}\right]$ and, correspondingly, $p:=-\imath \frac{\partial}{\partial q}= \left(\frac{-\imath}{\sqrt{e B}}\right) \frac{\partial}{\partial x_1}$, in terms of which the eigenvalue equation writes as
\begin{equation}\label{ME4}
    \begin{array}{c}\displaystyle
    \left\{{p}^2 + q^2 -\frac{\Lambda}{e B} \right\}\varphi(q)= 2 w
    \left\{ \left[ p +\imath q\right] + \imath  w \frac{E}{v_F B}\right\}\chi(q)\,,
       \\ \\ \displaystyle
       \left\{{p}^2 + q^2 -\frac{\Lambda}{e B} \right\}\chi(q)= 2 w
    \left\{ \left[ p -\imath q\right] - \imath  w \frac{E}{v_F B} \right\}\varphi(q)\,,
    \end{array}
\end{equation}
where
\begin{equation}\label{ME5}
    \Lambda:= {\lambda} + \frac{m E}{B}\left[ \frac{m E}{B} - 2 k \right]
\end{equation}
with $\lambda = {2 m} \mathcal{E}$.

\smallskip

For small electric fields, the last terms in the right hand sides of Eq.\ (\ref{ME4}) can be treated as the perturbation $\left(\frac{w^2 E}{m v_F B}\right) \sigma_2$ on a zero order Hamiltonian whose solutions are the same functions of $q$ as in Section \ref{Constant-magnetic}, with the equilibrium position of the oscillator in the $x_1$-direction displaced by an amount $\left(-\frac{mE}{e B^2}\right)$. The zero-order energy eigenvalues are then those of Section \ref{negative-mass} but shifted by an amount $\frac{E}{B}\left[k- \frac{m E}{2 B}\right]$.

\smallskip

On the other hand, since the perturbation is proportional to $\sigma_2$, it has a vanishing mean value due to the orthogonality of Hermite functions (See Eq.\ (\ref{B14})). Then, the spectrum is just rigidly shifted by the presence of an electric field perpendicular to the magnetic field, up to first order in perturbation theory in the small parameter $(E/v_F B)$.

\section{Conclusions and outlook}\label{Conclusions}

In this article we have studied a simple nonrelativistic model, suggested by a non-Abelian magnetic quantization approach (based on a Heisenberg algebra deformed by the introduction of an external  constant non-Abelian magnetic field), to describe the low energy excitations of graphene. As previously mentioned, these states admit an effective description as excitations of a (2+1)-dimensional pseudo-relativistic field theory of massless Dirac fermions.

Our effective model is based on a deformation of the Heisenberg algebra which makes the commutator of momenta proportional to the pseudo-spin. This can be interpreted as the commutator of two $U(2)$ covariant derivatives in the case of  a constant background non-Abelian magnetic field.  The modification introduced in the Heisenberg algebra leads to wave functions of two components and a  Hamiltoninan operator in the form of a $2\times 2$ matrix, which reduces to  the usually considered linear one in the large mass limit.

We have solved the Landau problem for this planar system, with a constant and uniform magnetic field applied perpendicularly to the plane, explicitly obtaining the Hamiltonian (generalized) eigenfunctions and eigenvalues. We have pointed out that this construction can be directly extended to the case of negative values of the mass parameter. This is important since a negative mass allows to reproduce with our Hamiltonian the low energy expansion of the dispersion relation around the first Dirac point of graphene, $\mathbf{K}$, when the leading order terms from both nearest and next-to-nearest neighbor interactions (linear and quadratic terms in the quasi-momentum expansion, respectively) are retained.

The effective Hamiltonian for the excitations around the second Fermi point, $H_{\mathbf{K}'}$, is proposed to be obtained from $H_{\mathbf{K}}$ through a transformation which maintains the Hamiltonian time-reversal times parity invariant.

Through this transformation, the spectrum of the low energy states for the second Dirac point $\mathbf{K'}$ is simply obtained from that of $\mathbf{K}$ by a reflection around the origin. This leads to an almost doubly degenerate spectrum for quasi-particles and holes, where the degeneracy is broken by $O\left(w^{-1}\right)$ terms, with $w=\frac{|m| v_F}{\sqrt{e B}}\simeq {10}^{3}$ for realistic values of the parameters. Moreover, there is one quasi-particle state of energy $\mathcal{E'}_0=eB/2|m|$ and one hole state of energy $\mathcal{E}_0=-eB/2|m|$.

The $|m| \rightarrow \infty$ limit reproduces the well known doubly degenerate spectrum of massless Dirac fermions, with one quasi-particle zero mode and one hole zero mode.

The Hall conductivity of the model has been evaluated from the partition function employing the $\zeta$-function approach to the associated functional determinant. This led to a rather general expression valid for the Landau problem of a fermionic theory. In the case under study, the result is consistent with the anomalous integer quantum Hall effect found in graphene, in which the (reduced) Hall conductivity as a function of the Fermi energy takes half-integer values (See Figure 3). In particular, $\frac{h}{4 e^2}\, \sigma_{xy}=\pm \frac{1}{2}$ for small (but $|\epsilon_F|> \frac{v_F\sqrt{eB}}{2w}$) positive or negative Fermi level, respectively.

Moreover, we have proposed an additional rotational symmetry breaking term, to be treated as a perturbation on the Hamiltonian, which reproduces the next-to-leading (quadratic) term in the nearest neighbor interaction contribution to the dispersion relation with no external field. With the perpendicular magnetic field present, we have shown that this term does not modify the spectrum at first order in perturbation theory. So, it neither changes the behavior of the Hall conductivity described above.

Finally, we have shown that an electric field perpendicular to the magnetic field produces just a rigid shift of the spectrum at first order in perturbation theory.

\medskip

An interesting problem under study refers to the case in which an Aharonov-Bohm singular flux is also present, which can simulate a defect in the lattice. This will be the subject of a forthcoming paper.

\vspace{0.4 cm}

\noindent{\textbf{Acknowledgements}}:  This  work was supported in part by grants from CONICET (PIP 01787), ANPCyT (PICT 00909) and UNLP (Proy.~11/X492), Argentina, and from FONDECYT (grants 1095217 and 1095106) and Proyecto Anillos ACT119, Chile. M.N. also acknowledge support from Universidad Nacional de La Plata, Argentina. We acknowledge E.M.\ Santangelo and C.G.\ Beneventano for useful discussions.
\smallskip

\appendix


\section{Lagrangian, conserved current and generating functional}\label{Ec-Mov}

In this Appendix we construct the Lagrangian for our model and deduce the conserved current associated with its $U(1)$  gauge symmetry. We also discuss the behavior of the generating functional in relation with the Hall conductivity.

The system under consideration can be described by the Lagrangian
\begin{equation}\label{L1}
\begin{array}{c}\displaystyle
      \mathcal{L}:=\frac{\imath}{2}\left[\psi^\dagger \, \partial_t \psi - \partial_t\psi^\dagger\, \psi\right]+
      \psi^\dagger e A_0 \psi
 \\ \\ \displaystyle
  -\frac{1}{2m} \left\{ \left[\left(\mathbf{p}-e\mathbf{A}+\theta\BS\right)\psi\right]^\dagger \cdot \left[\left(\mathbf{p}-e\mathbf{A}+\theta\BS\right)\psi\right]-2\theta^2\psi^\dagger\psi\right\}
\,.
\end{array}
\end{equation}

The variation of the independent dynamical variables $\psi^\dagger$ and $\psi$ leads to the Euler-Lagrange equations in the usual way:
\begin{equation}\label{L2}
    \begin{array}{c} \displaystyle
    \frac{\partial \mathcal{L}}{\partial \psi^\dagger}-\partial_t \left( \frac{\partial\mathcal{L}}{\partial \left(\partial_t \psi^\dagger\right)}\right) -
    \Bgrad\cdot \left( \frac{\partial\mathcal{L}}{\partial \left(\Bgrad \psi^\dagger\right)}\right)=
      \\ \\ \displaystyle
      =\imath \partial_t \psi- \frac{1}{2m} \left[\left(\mathbf{p}-e\mathbf{A}+\theta\BS\right)^2
      -2\theta^2\right] \psi =0\,,
    \end{array}
\end{equation}
and similarly for $\psi$.

\medskip

The Lagrangian in Eq.\ (\ref{L1}) has a $U(1)$ gauge symmetry. Indeed, it remains invariant against the following local change in the dynamical variables and the electromagnetic field:
\begin{equation}\label{L4}
    \begin{array}{c} \displaystyle
    \psi(x)\rightarrow e^{\imath e \alpha(x)}\psi(x) \quad \Rightarrow \quad  \delta \psi(x) = \imath e \alpha(x) \psi(x)\,,
      \\ \\ \displaystyle
      \psi^\dagger(x)\rightarrow \psi^\dagger(x) e^{-\imath e \alpha(x)} \quad \Rightarrow \quad  \delta \psi^\dagger(x)
       = -\imath e \alpha(x) \psi^\dagger(x)\,,
       \\ \\ \displaystyle
        A_\mu(x) \rightarrow A_\mu(x) + \partial_\mu \alpha(x)\,.
    \end{array}
\end{equation}
Then, the N{\oe}ther's Theorem implies the existence of a locally conserved current given by
\begin{equation}\label{L5}
    \alpha j^\mu := - \delta \psi^\dagger \left( \frac{\partial\mathcal{L}}{\partial \left(\partial_\mu \psi^\dagger\right)}\right)
    - \left( \frac{\partial\mathcal{L}}{\partial \left(\partial_\mu \psi\right)}\right) \delta\psi\,.
\end{equation}
Therefore, the charge density is
\begin{equation}\label{L6}
    j^0 = e\, \psi^\dagger \psi
\end{equation}
and the current density
\begin{equation}\label{L7}
    \mathbf{j}=\frac{e}{2m}\left\{
    \imath \left(\Bgrad \psi^\dagger\,  \psi - \psi^\dagger \, \Bgrad\psi \right)
    + 2 \psi^\dagger\left(-e\mathbf{A}+\theta\BS\right)\psi\right\}\,.
\end{equation}

It is a simple exercise to verify that $j^{\mu}$ is conserved as a consequence of the equations of motion, Eq.\ (\ref{L2}) and its complex conjugate,
\begin{equation}\label{L8}
    \partial_t j^0 - \Bgrad \cdot \mathbf{j}=0\,.
\end{equation}

\medskip

The generating functional of (current) Green's functions of this two-dimensional system reads as
\begin{equation}\label{L9}
    Z[A]=e^{\imath \Gamma[A]}:=\int \mathcal{D} \psi^\dagger \mathcal{D}\psi \, e^{\displaystyle
    \imath \int dt \int d^2 x \mathcal{L}(x)}\,,
\end{equation}
where $\Gamma[A]$ is the effective action. In particular, we have
\begin{equation}\label{L10}
    -\imath \frac{\delta\log Z[A]}{\delta A_\mu(x)}=\left\langle j^\mu(x) \right\rangle = J^\mu(x)
\end{equation}
and the second functional derivative gives minus the current-current correlation function.

\medskip

For a system with a gap in its spectrum (as is the case of our model with filled Landau levels), the effective action is a local gauge invariant functional of the electromagnetic external field \cite{Fradkin,Wen},
\begin{equation}\label{L11}
    \Gamma[A]=\int dt \int d^2 x \, \mathcal{L}_{eff}\left(A_\mu(x),\partial_\nu A_\mu(x),\dots\right)\,.
\end{equation}
In general, in an asymptotic weak field and weak gradient expansion, gauge invariance determines that the effective Lagrangian $\mathcal{L}_{eff}$ appears as an expansion in terms of the field intensities $\mathbf{E}$ and $B$ and its derivatives. But in 2+1 dimensions there is an additional possibility, since a Chern-Simons term can be the dominant one in this expansion (See, for example, \cite{Fradkin,Wen} and references therein),
\begin{equation}\label{L12}
    \mathcal{L}_{eff}= \frac{K}{4\pi}\, \epsilon^{\mu \nu \lambda} A_\mu \partial_\nu A_\lambda -
    \frac{1}{2}\, E_i \rho^{i j} E_j - \frac{1}{2} \, \chi B^2 + \dots\,,
\end{equation}
where $\epsilon^{\mu\nu\lambda}$ is the completely anti-symmetric tensor with $\epsilon^{012}=1$ and the ellipses represent higher order terms.

In fact, it is the Chern-Simons term which determines the Hall conductivity since
\begin{equation}\label{L13}
    J^\mu(x)= \frac{\delta \Gamma[A]}{\delta A_\mu(x)}=  \frac{K}{2\pi}\, \epsilon^{\mu \nu \lambda}  \partial_\nu A_\lambda +\dots = \frac{K}{4\pi}\, \epsilon^{\mu \nu \lambda}  F_{\nu \lambda} +\dots \,.
\end{equation}
In particular, for the spatial and temporal components of the current the previous expression reduces to
\begin{equation}\label{L14}
    J_i(x)=\frac{K}{2\pi}\, \epsilon_{0ij} E_j +\dots\,,
    \quad  J_0(x)=\frac{K}{2\pi}\, B + \dots\,.
\end{equation}
The first equation in \ref{L14} implies that the Hall conductivity $\sigma_{xy}=\frac{K}{2\pi}$ (or $\sigma_{xy}={K e^2}/{h}$ in full units).
Then,
\begin{equation}\label{L15}
    J_0=\sigma_{xy} B\,.
\end{equation}

\medskip

It can be shown that $K$ takes integer values if the (many-body) ground state is not degenerate and has a finite energy gap \cite{Fradkin,Wen,Bellissard}. In this case, $K$ has a topological character: It can be related to a Berry phase which, in turns, takes the value $2\pi$ times an (integer) Chern number.


\begin{thebibliography}{99}


\bibitem{1}{L.D. Landau, Phys. Z. Sowjet Union 11 (1937) 26.}

\bibitem{2}{P.W. Anderson, Basic Notions of Condensed Matter Physics, Addison-Wesley, Reading, MA, 1997.}

\bibitem{3}{See {\it e.g}, C. Itzikson and J. B. Zuber, Quantum Field Theory, McGraw-Hill, 1980}

\bibitem{4}{A. Linde, Particle Physics and Inflationary Cosmology, Harwood Acad, Chur, Switzerland, 1990.}

\bibitem{5}{J.C. Collins, Renormalization, Cambridge Univ. Press, Cambridge, 1984.}

\bibitem{6}{K.G. Wilson, J. Kogut, Phys. Rep. 12 (1974) 75.}

\bibitem{7}{S.K. Ma, Modern Theory of Critical Phenomena, 1976.}

\bibitem{8}{K.G. Wilson, Rev. Modern Phys. 47 (1975) 773.}

\bibitem{9}{A.C. Hewson, The Kondo Problem to Heavy Fermions, Cambridge University Press, Cambridge, 1993.}


\bibitem{Novo-2004}{Novoselov, K. S., A. K. Geim, S. V. Morozov, D. Jiang, Y. Zhang, S. V. Dubonos, I. V. Gregorieva, and A. A. Firsov, 2004, Science 306, 666.}

\bibitem{Castro}{A. H. Castro Neto, F. Guinea, N. M. R. Peres, K. S. Novoselov and A. K. Geim,  Rev.\ Mod.\ Phys.\ Vol.\ 81, (2009) 109.}

\bibitem{Vozmediano}{M.A.H.\ Vozmediano, M.I.\ Katsnelson and F.\ Guinea, Physics Reports 496, (2010) 109–148.}

\bibitem{Semenoff}{G.W.\ Semenoff, Phys.\ Rev.\ Lett.\ 53, (1984) 2449.}

\bibitem{Wallace}{Wallace, P. R., 1947, Phys. Rev. 71, 622.}

\bibitem{Gusynin-2005}{Gusynin, V. P., and S. G. Sharapov, 2005, Phys. Rev. Lett. 95, 146801.}

\bibitem{Peres-2006}{Peres, N. M. R., F. Guinea, and A. H. Castro Neto, 2006a, Phys. Rev. B 73, 125411.}

\bibitem{Novo-2005}{Novoselov, K. S., A. K. Geim, S. V. Morozov, D. Jiang, M. I. Katsnelson, I. V. Grigorieva, S. V. Dubonos, and A. A. Firsov, 2005, Nature 438, 197.}

\bibitem{Zhang}{Zhang, Y., Y.-W. Tan, H. L. Stormer, and P. Kim, 2005, Nature 438, 201.}

\bibitem{Snyder}{H.S. Snyder, Phys. Rev. 71 (1947) 38; H.S. Snyder, Phys. Rev. 72 (1947) 68.}

\bibitem{Landau}{See L.D. Landau, E.M. Lifshitz, M\'{e}canique Quantique, \'{E}ditions Mir, 1966, p. 496.}

\bibitem{Douglas}{The literature is vast in this field, for a general review see M.R. Douglas, N.A. Nekrasov, Rev. Mod. Phys. 73 (2001) 977.}

\bibitem{Connes1}{A. Connes, "Noncommutative geometry," Academic Press, London and San Diego (1994).}

\bibitem{Connes2}{A. Connes and M. Marcolli, Colloquium
Publications, Vol. 55, American Mathematical Society (2008).}

\bibitem{Witten}{E. Witten,  Nucl. Phys. B 268, 253 (1986).}

\bibitem{Seiberg}{N. Seiberg and E. Witten,  JHEP 9909, 032 (1999).}

\bibitem{Hinchliffe}{I. Hinchliffe, N. Kersting, Y.L. Ma, Int.J. Mod. Phys. A 19 (2004) 179.}



\bibitem{Doplicher}{S.\ Doplicher, K.\ Fredenhagen, J.E.\ Roberts, Comm.\ Math.\ Phys.\ 172 (3), 187220 (1995).}

\bibitem{Rie}{M.A.\ Rieffel, Proc.\ Symp.\ Pure Math., 51, AMS, Providence, 411-423, 1990.}

\bibitem{Ift}{V.\ Iftimie, M.\ Mantoiu, R.\ Purice, Publ.\ RIMS.\ 43 (3), 585-623 (2007).}

\bibitem{Man}{M.\ Mantoiu, R.\ Purice, J.\ Math.\ Phys.\ 46, 052105 (2005).}


\bibitem{NCQM}{The literature is very extensive; a partial list can be found in the references of \cite{SpinNC2}}

\bibitem{Dayi}{Omer F. Dayi, Ahmed Jellal, J.~Math.~Phys. 51 (2010) 063522.}

\bibitem{SpinNC1}{H. Falomir, J. Gamboa, J. Lopez-Sarrion, F. M\'{e}ndez and P. Pisani, Phys. Lett., 680, 384 (2009).}

\bibitem{SpinNC2}{Ashok Das, H. Falomir, M. Nieto, J. Gamboa, F. Mendez, Phys. Rev. D84 (2011) 045002.}

\bibitem{weiss} {L.S. Brown and W.I. Weisberger, Nucl.Phys. B157 (1979) 285.}

\bibitem{Mariel}{C G Beneventano, Paola Giacconi, E M Santangelo and Roberto Soldati,  J.\ Phys.\ A: Math.\ Theor.\ 42 275401 (2009).}

\bibitem{McClure}{J. W. McClure, 1956, Phys. Rev. 104, 666.}

\bibitem{AS}{M.F. Atiyah and I.M. Singer, Ann. of Math. 87 (1968) 484. Proc. Natl. Acad, Sci. USA 81 (1984) 2597.}

\bibitem{rede}{A. N.  Redlich, Phys. Rev. D 29 (1984) 2369.}


\bibitem{Fradkin}{E.\ Fradkin, "Field Theory of Condensed Matter Systems",
Addison-Wesley Publishing Company  (1991), Redwood City, California, USA.}

\bibitem{Wen}{Xiao-Gang Wen, "Quantum Field Theory of Many-Body Sistems", Oxford University Press (2004), Oxford, Great Britain.}

\bibitem{Dow}{J.S.\ Dowker and R.\ Critchley, Phys. Rev. D13, 3224 (1976).}

\bibitem{Seeley}{R.T.\ Seeley, A.M.S.\ Proc.\ Symp.\ Pure Math. 10, 288 (1967). Am.\ Journ.\ Math.\ 91, 889
(1969). Am.\ Journ.\ Math.\ 91, 963 (1969).}


\bibitem{Mathematica}{Mathematica 8.0, Wolfram Research Inc. (2011).}

\bibitem{MathWorld}{Sondow, Jonathan and Weisstein, Eric W, "Hurwitz Zeta Function." From MathWorld--A Wolfram Web Resource. http://mathworld.wolfram.com/HurwitzZetaFunction.html}

\bibitem{Lukose}{V. Lukose, R. Shankar, and G. Baskaran, 2007, Phys. Rev. Lett. 98, 116802.}

\bibitem{Castro-Peres}{N. M. R. Peres and E. V. Castro, 2007, J. Phys.: Condens. Matter 19, 406231.}

\bibitem{Mariel2}{C G Beneventano, Paola Giacconi, E M Santangelo and Roberto Soldati, }

\bibitem{Bellissard}{J. Bellissard, A. van Elst, and H. Schulz-Baldes, J.\ Math.\ Phys.\ 35, 5373 (1994).}










\end{thebibliography}
\end{document}